Comment on "Evolution of the electronic excitation spectrum with strongly diminishing hole density in superconducting $Bi_2Sr_2CaCu_2O_{8+x}$" (J. W. Alldredge *et al.*, *Nature Phys.* **4**, 319 (2008))

In a recent article, Alldredge *et al.*[1] reported tunnelling data obtained at $T$ = 4.2 K by STM in $Bi_2Pb_2Sr_2Cu_2O_{8+x}$ (Bi-2212) having different doping levels, $p$, and a new way to fit the measured differential conductances, $g(V)$, which are linearly proportional to the local density of states of quasiparticle excitations at the surface, $N(E)$. The main point of the fitting procedure is to use an energy-dependent inelastic scattering rate, $\Gamma = \alpha E$ ($\alpha$ is a constant). They argue that by using an equation of the *d*-wave BCS density of states[2] (Bardeen-Cooper-Schrieffer) modified by Dynes *et al.*[3] with $\Gamma = \alpha E$, they were able to fit any conductance obtained in Bi-2212 with $0.1 \leq p \leq 0.22$. In addition, they discuss a new energy scale $\Delta_0(p)$ similar to that obtained in Raman measurements[4]. Unfortunately, the paper is misleading.

The presentation of experimental data in the article is excellent; however, the authors ignored the fact that cuprates are not BCS-type[2] superconductors (SCs). Therefore, an attempt to use the BCS equation, modified or not, to fit tunnelling conductances obtained in cuprates is erroneous. Even, a simple visual inspection of some of their graphs shows a poor correspondence between measured data and the modified BCS function (see Fig. S3). In particular, the fit reproduces poorly measured conductances outside the gap structure (dips, humps), i.e. at high bias (this is the reason why all the spectra are presented **only** between ±90 mV), and inside the gap structure. The subgap kinks which will be discussed below are completely disregarded by the fit, as also admitted by the authors. **If** the authors kept in mind that cuprates are not BCS-type SCs, and they were just looking for a suitable function to fit the data, they should write a few sentences to explain their approach to the problem. This has not been done. In any case, the fit which they propose to use, unfortunately, is not good enough to reproduce all the features of tunnelling conductances in Bi-2212.

Concerning the physics, in photoemission measurements in cuprates, as the temperature decreases,

the quasiparticle peak appear in the spectra at the inner side of hump[5,6], thus, inside of a normal-state gap. The same happens in tunnelling measurements in cuprates: quasiparticle peaks appear inside the humps[7-11]. This means that the "shoulders" (humps) of tunnelling conductances obtained in cuprates and quasiparticle peaks correspond to different electronic states in cuprates (for example, charge-density waves (CDW) and SC, respectively). Moreover, cuprates have two energy gaps[8,9,12] which are both involved somehow in high-$T_c$ SC[8,9]. So, to fit successfully tunnelling conductances of cuprates, we still have to search for an appropriate fit. This problem is directly related to the understanding of the mechanism of high-$T_c$ SC. The fit should be applicable along the whole SC dome ($0.05 \leq p \leq 0.27$).

Returning to the article, the idea of an inelastic scattering rate in cuprates linearly dependent on energy is not new (see, e.g. ref. 13). Apart from the extended BCS fit, the authors discuss a new energy scale, $\Delta_0(p)$, based on subgap kinks in tunnelling conductances. Indeed, tunnelling conductances of cuprates often have subgaps, and their temperature dependence is different from that of quasiparticle peaks[8]. The kinks may have a relation with a smaller, "homogeneous" SC gap seen in Bi-2201[12].

As it concerns the text, for an article in Nature Physics, the text could be written better: (i) in the abstract, the first three sentences have no logical connection with the rest starting with "Here, we ...". (ii) On page 320, the authors use an expression "two 'coherence' peaks in $g(V)$". Tunnelling spectroscopy is phase-insensitive. Conductance peaks are caused by quasiparticle excitations, and not by long-range phase coherence. (iii) The equation (2) contains the letters "$f_s$" (or "$fs$") without an explanation in the text a meaning of it. (iv) Throughout the text, the doping level, $p$, is expressed in some places in %, in some in absolute value (on page 321, even in adjacent sentences!). (v) The expressions in the text "we introduce a new technique" and "another breakthrough" are overstated. The efforts which the authors did are enormous; however, the goal is not achieved.

I acknowledge helpful communications with E. Hudson.

A. Mourachkine

Cavendish Laboratory, University of Cambridge, J.J. Thomson Ave., Cambridge CB3 0HE, UK